\documentclass[5p,twocolumn]{elsarticle}

\usepackage{lineno,hyperref}
\modulolinenumbers[5]

\usepackage{amsmath}
\usepackage{amssymb}

\usepackage{graphicx}
\usepackage{float}
\usepackage{xcolor}
\usepackage[english]{babel}
\usepackage[output-decimal-marker={.},exponent-product=\cdot]{siunitx}

\biboptions{sort&compress}

\hypersetup{colorlinks,allcolors=black}
\journal{The European Physical Journal A}

\begin{document}

\begin{frontmatter}

\title{Precision mass measurement of $^{173}$Hf for nuclear structure of $^{173}$Lu and the $\gamma$ process}

\author[a]{A.~Jaries\corref{cor}}
\ead{arthur.a.jaries@jyu.fi}
\author[a]{M.~Stryjczyk\corref{cor}}
\ead{marek.m.stryjczyk@jyu.fi}
\author[a]{A.~Kankainen\corref{cor}}
\ead{anu.kankainen@jyu.fi}
\author[a]{T.~Eronen}
\author[a,b]{Z.~Ge}
\author[a,c]{M.~Hukkanen}
\author[a]{I.D.~Moore}
\author[a]{M.~Mougeot}
\author[a]{A.~Raggio}
\author[a]{W.~Rattanasakuldilok}
\author[a]{J.~Ruotsalainen}

\address[a]{University of Jyvaskyla, Department of Physics, Accelerator laboratory, P.O. Box 35(YFL) FI-40014 University of Jyvaskyla, Finland}
\address[b]{GSI Helmholtzzentrum f\"ur Schwerionenforschung, 64291 Darmstadt, Germany}
\address[c]{Universit\'e de Bordeaux, CNRS/IN2P3, LP2I Bordeaux, UMR 5797, F-33170 Gradignan, France}
\cortext[cor]{Corresponding authors}

\begin{abstract}
We report on the precise mass measurement of the $^{173}$Hf isotope performed at the Ion Guide Isotope Separator On-Line facility using the JYFLTRAP double Penning trap mass spectrometer. The new mass-excess value, ${\mathrm{ME} = -55390.8(30)}$~keV, is in agreement with the literature while being nine times more precise. The newly determined $^{173}$Hf electron-capture $Q$ value, $Q_{EC} = 1490.2(34)$~keV, allows us to firmly reject the population of an excited state at 1578 keV in $^{173}$Lu and 11 transitions tentatively assigned to the decay of $^{173}$Hf. Our refined mass value of $^{173}$Hf reduces mass-related uncertainties in the reaction rate of $^{174}$Hf$(\gamma,n)^{173}$Hf. Thus, the rate for the main photodisintegration destruction channel of the $p$ nuclide $^{174}$Hf in the relevant temperature region for the $\gamma$ process is better constrained. 
\end{abstract}

\begin{keyword}
\texttt Binding energies and masses \sep Mass spectrometers \sep Penning trap
\end{keyword}

\end{frontmatter}


\section{Introduction}

The radioactive neutron-deficient $^{173}_{72}$Hf$_{101}$ isotope (${T_{1/2} = 23.6(1)}$~h \cite{NUBASE2020}) was identified for the first time in 1951 \cite{Wilkinson1951}. Since then, its decay to $^{173}$Lu was studied in several experiments, see Ref. \cite{Shirley1995} and references therein. The current knowledge of the $^{173}$Hf decay is based primarily on the two most recent studies, reported by Funk \textit{et al.} \cite{Funk1974} and by Brenner \textit{et al.} \cite{Brenner1975}. In both of these works the radioactive isotope of interest was produced by irradiating enriched Yb targets ($^{172,173}$Yb in Ref.~\cite{Funk1974} and $^{172}$Yb in Ref.~\cite{Brenner1975}) with an $\alpha$ beam. The isotope of interest was extracted by means of chemical separation and the $\gamma$-ray radiation following the decay was measured using Ge(Li) detectors. In addition, in Ref. \cite{Funk1974} the conversion electrons were detected using a Si(Li) detector. 

At the time of $^{173}$Hf decay studies publication \cite{Funk1974,Brenner1975}, the electron-capture $Q$ value ($Q_{EC}$) estimated from the systematics was 1600~keV \cite{Wapstra1971}. Nonetheless, several $\gamma$-ray and conversion-electron transitions with an energy above 1.6~MeV were tentatively assigned to its decay \cite{Funk1974,Brenner1975}. To resolve this disagreement Funk \textit{et al.} assumed that the $Q_{EC}$ value is 1900~keV \cite{Funk1974}. On the other hand, Brenner \textit{et al.} indicated that they assigned weak high-energy $\gamma$ rays to the decay of $^{173}$Hf when these transitions could not be associated with known impurities \cite{Brenner1975}. 

A mass measurement of $^{173}$Hf performed at the GSI storage ring \cite{Litvinov2005} and the resulting decrease of $Q_{EC}$ to 1469(28)~keV \cite{AME2020} rendered the decay spectroscopy results incompatible. In addition to the high-energy transitions, a 1578-keV excited state in $^{173}$Lu proposed in the work by Funk \textit{et al.} \cite{Funk1974}, was also found to be inconsistent with the new $Q_{EC}$ value. An independent evaluation of the $^{173}$Hf mass and, consequently, of the $Q_{EC}$ value would enable a resolution of the aforementioned issues. Such a measurement would allow us to unambiguously establish whether the high-energy transitions were correctly assigned and verify the presence of the 1587-keV state. 

In addition to the nuclear spectroscopy interest, the mass of $^{173}$Hf is also of relevance for the astrophysical $p$ process, also known as the $\gamma$ process \cite{Woosley1978,Arnould2003}. The process proceeds mainly via photodisintegration reactions and takes place in thermonuclear and core-collapse supernovae, when the shock wave passes through the O-Ne layer at typical temperatures of around 1.5-4 GK. The $\gamma$ process produces altogether 35 stable isotopes. The production of heavier $p$-process isotopes, such as the long-lived radionuclide $^{174}$Hf ($T_{1/2}=2.0(4)\times 10^{15}$~y \cite{NUBASE2020}), is very sensitive to temperature \cite{Rapp2006}. This is due to the competition between two highly temperature-dependent reactions, namely, $(\gamma,\alpha)$ and $(\gamma,n)$. As the level densities are high for such heavy isotopes, the reaction rates are typically calculated using the statistical Hauser-Feshbach (HF) approach \cite{Hauser-Feshbach1952}. Precise knowledge of the corresponding  ground-state properties of the target nucleus and residual nuclei, such as masses, is needed for the HF calculations \cite{Arnould2003}. 

In this work, we constrain the mass-related uncertainties related to the $^{174}$Hf$(\gamma,n)^{173}$Hf reaction via a high-precision mass measurement of $^{173}$Hf at the JYFLTRAP double Penning trap. The results are discussed in the context of the nuclear structure and the astrophysical $\gamma$ process.

\section{Experimental method and results}

The experiment was performed at the Ion Guide Isotope Separator On-Line (IGISOL) facility \cite{Moore2013,Penttila2020} at the University of Jyv\"askyl\"a, Finland. Both the $^{173}$Hf isotope of interest and the $^{173}$Yb reference-mass isotope, were produced in a fusion-evaporation reaction of a 50-MeV $\alpha$ beam, delivered by the K130 cyclotron with an average current of 1.1~p$\mu$A, and a 1.75~mg/cm$^{2}$-thick $^{nat}$Yb target mounted within the light-ion ion guide. The reaction products were stopped in a helium-filled gas cell operating at about 250~mbar. The ions were subsequently extracted with gas flow and guided to the high-vacuum region of the mass separator using a sextupole ion guide \cite{Karvonen2008}, accelerated by a 30-kV potential and mass-separated by a 55$^{\circ}$ dipole magnet. The continuous beam was injected into the radio-frequency quadrupole cooler-buncher \cite{Nieminen2001} where it was cooled and bunched. From there the radioactive ion beam was finally delivered to the JYFLTRAP double Penning trap \cite{Eronen2012}. 

In JYFLTRAP, the singly-charged $A=173$ ions were first cooled, purified to contain only $^{173}$Yb, $^{173}$Lu and $^{173}$Hf, and centered using a mass-selective buffer gas cooling technique \cite{Savard1991} in the first trap. After that, the ions were sent to the second (measurement) trap where their charge-over-mass-dependent ($q/m$) cyclotron frequency $\nu_c = qB/(2 \pi m)$ in a magnetic field $B$ was measured by using a phase-imaging ion cyclotron resonance (PI-ICR) technique \cite{Eliseev2013,Eliseev2014,Nesterenko2018,Nesterenko2021}. 

\begin{figure}[h!t!b]
\includegraphics[width=\columnwidth]{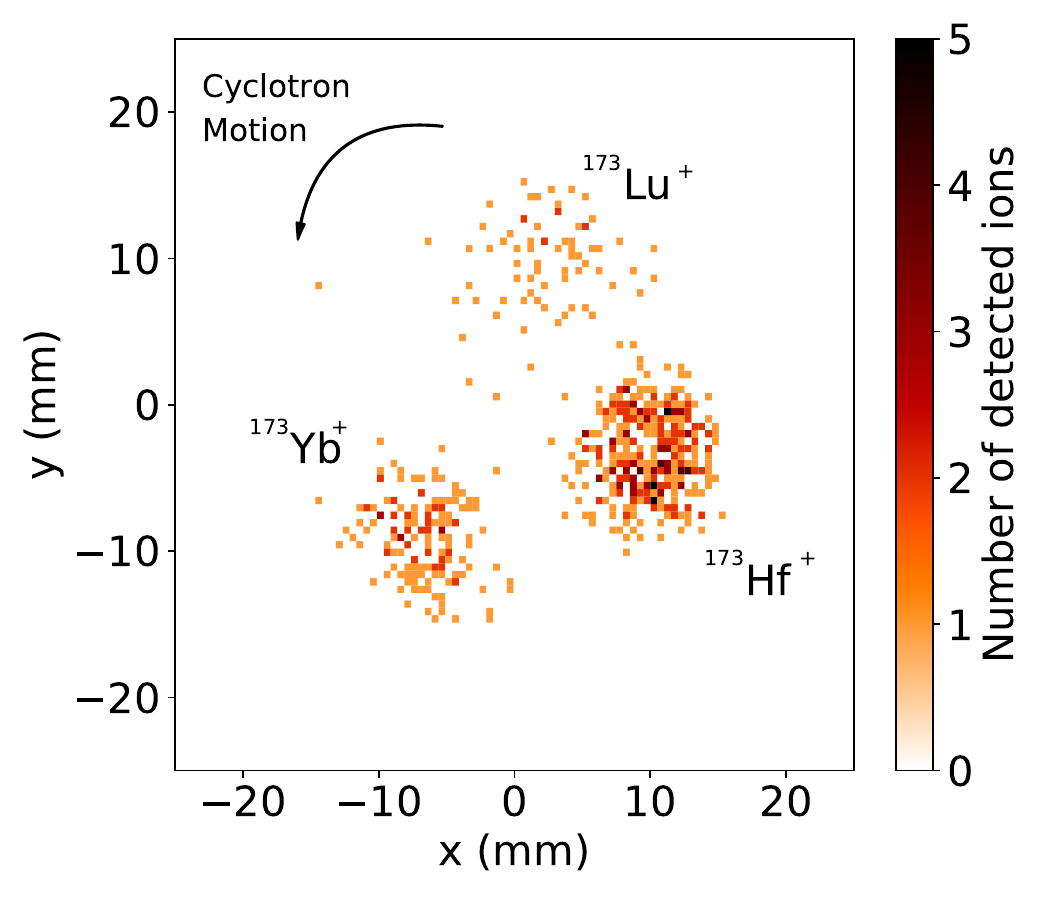}
\caption{\label{fig:173HfPIICR}Projection of the cyclotron motion of $^{173}$Hf$^+$ and the isobaric contaminants of $^{173}$Lu$^+$ and $^{173}$Yb$^+$ ions onto the position-sensitive detector obtained with the PI-ICR technique using a phase accumulation time $t_{acc} = 584$~ms. The total number of ions is 764 and the number of ions per bunch has not been limited for this figure.}
\end{figure}

Using the PI-ICR technique, the cyclotron frequency $\nu_c$ of an ion is obtained from the phase differences between its radial in-trap motions during a phase accumulation time $t_{acc}$ (see Fig.~\ref{fig:173HfPIICR}). In the present case, $t_{acc}$ value was set to 584~ms to avoid an overlap between the projections of $^{173}$Hf (ion of interest), $^{173}$Lu (isobaric contaminant) and $^{173}$Yb (reference mass). The final mass value for $^{173}$Hf is obtained by measuring a cyclotron frequency ratio between $^{173}$Hf and $^{173}$Yb. For the $^{173}$Yb reference, the mass excess reported in the Atomic Mass Evaluation 2020 (AME20), ${\mathrm{ME}_{lit.} = -57551.234(11)}$~keV \cite{AME2020} and based on the Penning trap measurement \cite{Rana2012} was used. The measurements of the ion of interest and the reference ion were alternated every $\sim$5 minutes to account for the temporal magnetic field fluctuations.

The energy difference between $^{173}$Yb and $^{173}$Hf isotopes, $\Delta E$, was calculated using the cyclotron frequency ratio ${r=\nu_{c,ref.}/\nu_{c}}$ of singly-charged ions of both species:
\begin{equation}
\Delta E = (r-1)[m_{ref} - m_e]c^2 \mathrm{,}
\end{equation} 
with $m_e$ and $m_{ref}$ being the masses of a free electron and the atomic mass of $^{173}$Yb, respectively, and $c$ being the speed of light in vacuum. The contribution from electron binding energies are on the order of a few eV and have thus been neglected. To reduce any systematic uncertainty due to ion-ion interactions, the count rate was limited to one detected ion per bunch. The systematic uncertainties due to the magnetron phase advancement, the angle error and the temporal magnetic field fluctuation $\delta B/B = 2.01(25) \times 10^{-12}$ min$^{-1} \times \delta t$ with $\delta t$ being the time between the measurements were taken into account \cite{Nesterenko2021}. However, their effect ($\delta r/r\sim6\times10^{-9}$, $\sim2\times10^{-9}$ and $\sim2\times10^{-11}$, respectively) is much smaller compared to the statistical uncertainty ($\delta r/r\sim2\times10^{-8}$).

\begin{table}
\centering
\caption{\label{tab:results} A comparison of the energy difference between $^{173}$Hf and $^{173}$Yb ($\Delta E$), the mass excess of $^{173}$Hf ($\mathrm{ME}(^{173}\mathrm{Hf})$) and its electron-capture $Q$ value ($Q_{EC}(^{173}\mathrm{Hf})$) between this work and AME20 \cite{AME2020}. The cyclotron frequency ratio $r=\nu_{c,ref}/\nu_{c}$ determined in this work using the PI-ICR technique is also reported.}
\begin{tabular}{llll}
\hline\noalign{\smallskip}
Quantity & AME20 & This work \\\hline
$r=\nu_{c,ref}/\nu_{c}$ & & \num{1.000013411(19)} \\
$\Delta E$ (keV) & \num{2139(28)} & \num{2160.4(30)} \\
$\mathrm{ME}(^{173}\mathrm{Hf})$ (keV) & \num{-55412(28)} & \num{-55390.8(30)} \\
$Q_{EC}(^{173}\mathrm{Hf})$ (keV) & $1469(28)$ & $1490.2(34)$ & \\\hline
\noalign{\smallskip}
\end{tabular}
\end{table}

The experimental results are summarized in Table~\ref{tab:results}. The calculated $\Delta E$ value and the deduced mass excess of $^{173}$Hf (${\mathrm{ME}(^{173}\mathrm{Hf}) = \mathrm{ME}(^{173}\mathrm{Yb}) + \Delta E}$) are in agreement with the literature (${\mathrm{ME} - \mathrm{ME}_{lit.} = 21(28)}$~keV), however, our result is nine times more precise. To obtain $Q_{EC}(^{173}\mathrm{Hf}) = \mathrm{ME}(^{173}\mathrm{Hf}) - \mathrm{ME}(^{173}\mathrm{Lu})$, the ${\mathrm{ME}_{lit.}(^{173}\mathrm{Lu}) = -56881.0(16)}$~keV was taken from AME20 \cite{AME2020} and it leads to ${Q_{EC}(^{173}\mathrm{Hf}) = 1490.2(34)}$~keV. The updated value agrees with the literature ($1469(28)$~keV \cite{AME2020}), however, it is eight times more precise. 

The agreement between the mass measurements reported in Ref. \cite{Litvinov2005} and this work allows us to unambiguously reject the hypothesis of the 1578-keV state being populated in the $\beta$ decay of $^{173}$Hf. We can also remove seven transitions (1505, 1551.0, 1557.7, 1749, 1778.4, 1836 and 1897 keV) assigned to the decay of $^{173}$Hf in Ref.~\cite{Funk1974} and four transitions (1512.5, 2043.0, 2127.7 and 2613.1 keV) from Ref.~\cite{Brenner1975}. We note that there are two transitions at 1485.1~keV \cite{Funk1974} and at 1488.9~keV \cite{Brenner1975} which are within 2$\sigma$ of the updated $Q_{EC}$ value, therefore, they cannot be unambiguously removed or kept in the $^{173}$Hf decay scheme. 

There are several possible explanations why the aforementioned transitions were incorrectly assigned to the decay of $^{173}$Hf. All of them have a very low absolute intensity ($I_\gamma < 10^{-3}$) which hindered the $\gamma-\gamma$ coincidence analysis. The radioactive samples were prepared using chemical separation methods which are known to have a limited reliability. As a result, transitions originating from different species could also be observed. In addition, the low $Q_{EC}$ value prevented the $\beta-\gamma$ coincidence analysis as the vast majority of the decays underwent the electron capture channel \cite{Shirley1995}. This could result in an accidental assignment of the background transitions to the decay scheme.

\begin{figure}[h!t!b]
\includegraphics[width=\columnwidth]{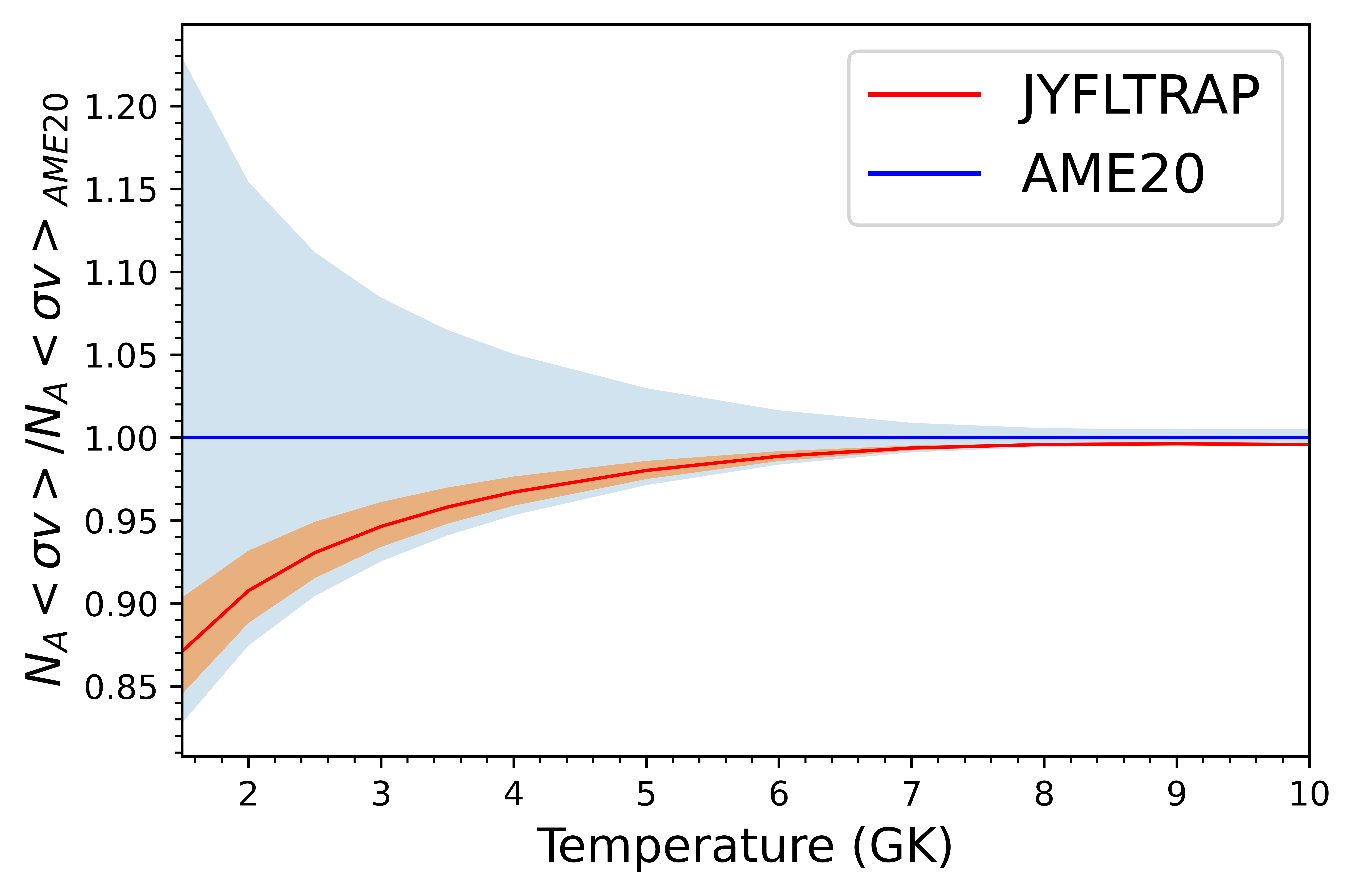}
\caption{\label{fig:gamman_rr}Astrophysical reaction rate ratio using the mass of $^{173}$Hf determined in this work (JYFLTRAP) and the AME20 mass values for the $^{174}$Hf($\gamma,n$)$^{173}$Hf reaction as a function of temperature. The $(\gamma,n)$ is the main photodisintegration destruction channel of $^{174}$Hf down to temperatures of 2~GK.}
\end{figure}

The more precise mass value of $^{173}$Hf is also relevant for constraining the calculated photodisintegration reaction rate on the $p$ nuclide $^{174}$Hf. Although the natural abundance of the radionuclide $^{174}$Hf is rather low, 0.16(1)\%, it has been shown that it can be used as a tracer to explore the distribution of supernova material in the early solar system  \cite{Peters2017}. Constraining the photodisintegration reaction rates of $^{174}$Hf has an impact not only on the $^{174}$Hf abundance but also on the lighter $p$ nuclide abundances as the process eventually proceeds to lighter elements via $(\gamma,p)$ and $(\gamma,\alpha)$ reactions. Here we constrain the $^{174}$Hf$(\gamma,n)^{173}$Hf reaction rate with the new, more precise mass value of $^{173}$Hf (see Fig.~\ref{fig:gamman_rr}). The astrophysical reaction rates were calculated with the TALYS-1.96 code \cite{talys}, using the default phenomenological level density  model based on the Fermi gas model and the local optical model potential parametrization \cite{talys}. The masses of $^{173}$Hf and $^{174}$Hf were adopted from this work (JYFLTRAP) and AME20 \cite{AME2020} and varied up or down by $1\sigma$ to obtain the maximum and minimum $Q$ values for the reaction of interest.

The updated $Q$ value for the $^{174}$Hf$(\gamma,n)^{173}$Hf reaction resulted in a reaction rate decrease by up to 13\% for the relevant temperature region compared to the rate calculated with the AME20 masses, see Fig. \ref{fig:gamman_rr}. The $(\gamma,n)$ reaction is the main photodisintegration destruction channel of $^{174}$Hf for temperatures down to 2 GK below which the $(\gamma,\alpha)$ starts to dominate. The total photodisintegration reaction rate, however, decreases significantly at those lower temperatures. Although there are also many other uncertainties related to the reaction rates and the $p$-nuclide abundances (see e.g. \cite{Nishimura2017,Rauscher2016}), the mass-related reaction rate uncertainties for the main destruction channel of the $p$ nuclide $^{174}$Hf were significantly reduced in this work, e.g. from $\approx$14\% to $\approx$2.4\% at 2.0 GK. 

\section{Conclusions}

The mass of $^{173}$Hf was measured with high precision using the PI-ICR method at the JYFLTRAP double Penning trap. The result is in agreement with the literature data, however, it is nine times more precise. The updated $Q_{EC}$ value of $^{173}$Hf allowed us to exclude one excited state and 11 transitions in the daughter nucleus $^{173}$Lu, previously assigned to the decay $^{173}$Hf. The high-precision mass measurement also constrained the calculated $(\gamma,n)$ photodisintegration rate on the $p$ nucleus $^{174}$Hf.

\section*{Acknowledgments}

This project has received funding from the European Union’s Horizon 2020 research and innovation programme under grant agreements No. 771036 (ERC CoG MAIDEN) and No. 861198–LISA–H2020-MSCA-ITN-2019 and from the Academy of Finland projects No. 295207, 306980, 327629, 354589 and 354968. J.R. acknowledges financial support from the Vilho, Yrj\"o and Kalle V\"ais\"al\"a Foundation.

\bibliography{mybibfile}

\end{document}